**Title**: OWL-Moon: Very high resolution spectropolarimetric interferometry and imaging from the Moon: exoplanets to cosmology

Jean Schneider, Observatoire de Paris, France
jean.schneider@obspm.fr

Joseph Silk, IAP (France)/John Hopkins University (USA)
joseph.silk@physics.ox.ac.uk

Farrokh Vakili, Observatoire de la Côte d'Azur, Nice, France
farrokh.vakili@oca.eu

**Abstract**
We outline a concept for OWL-Moon, a 50-100m aperture telescope located on the surface of the Moon, to address three major areas in astronomy, namely the detection of biosignatures on habitable exoplanets, the geophysics of exoplanets, and cosmology. Such a large lunar telescope, when coupled with large Earth-based telescopes, would allow Intensity Interferometric measurements, leading to pico-arcsecond angular resolution. This would have applications in many areas of astronomy and is timely in light of the renewed interest of space agencies in returning to the Moon.

**Introduction**
The search for life on exoplanets is one of humanity's greatest challenges. The focus for the next decade will be the search for super-Earths, revealed by the CoRoT and Kepler missions to predominate among exoplanets in our interstellar neighborhood. Detecting super-Earths, some of which may support liquid water oceans, requires larger and more ambitious telescopes than those currently in operation.

One concept under study for the 2030s is the LUVOIR mission (LUVOIR Team, 2019), an 8-15m aperture telescope to be launched to L2, with imaging and spectroscopic capability spanning the ultraviolet to near-infrared (200-2000 nm). LUVOIR will address a broad range of science goals including the characterization of habitable-zone exoplanets and the search for biosignatures.

In the context of ESA's Voyage 2050 long-term plan for the ESA science programme, we demonstrate that, following on from LUVOIR, a large telescope located on the

Moon can perhaps provide the first serious opportunity within the next few decades for assessing the complexity of extraterrestrial life on Earth-like exoplanets.

Tackling these science goals requires a very large aperture to detect spectro-polarimetric and spatial features, from the UV to near-infrared, of faint objects such as exoplanets, and continuous monitoring to characterize the temporal behavior of exoplanets, such as rotation period, meteorology, and seasons. In addition to exoplanets, an OWL-Moon will have several applications in cosmology, stellar physics, and other areas (see below).

Terrestrial telescopes are not suited for continuous monitoring of celestial sources and the atmosphere limits the ultimate angular resolution and spectro-polarimetric domain that can be attained. Since the aperture of telescopes in orbit is limited to perhaps 15m over the next several decades, the need for a lunar platform for an Overwhelmingly Large Telescope (OWL) based on ESO's OWL concept is apparent.

In addition, by applying a novel approach using Intensity Interferometry (II) with Earth-based telescopes, it will be possible to achieve angular resolution of the order of pico-arcseconds. Further details of the OWL-Moon concept and the application of Intensity Interferometry can be found in an extended version (Schneider et al. 2021) of this short communication, which was submitted as a White Paper to the Voyage 2050 Call (Schneider et al, 2019).

**Science objectives**

We outline here the major topics an OWL-Moon will address:

*1/ Exoplanets*
First, with its large, 50m or more, aperture, OWL-Moon will characterize the atmosphere of super-Earths and provide hints for biosignatures. Another, original, capability of OWL-Moon, taking advantage of the unprecedented angular resolution provided by Earth-Moon Intensity Interferometry, will be the possibility of measuring the height of mountains of transiting planets.

*2/ Stellar Physics*
Examples of major contributions to our understanding of stars from ultrahigh-resolution spectropolarimetry include: studies of poorly understood atmospheric dynamics, including starspots; the role of magnetic fields in the evolution of different types of stars; and the launching of stellar winds from massive stars. Spectropolarimetry will enable us to probe stellar environments, ranging from the dynamic upper atmospheres of cool stars to their circumstellar discs and planetary systems. Asteroseismology uniquely explores stellar interiors, and high resolution spectropolarimetry will greatly facilitate this field. Solar seismology is in strong tension with the standard solar model, and it is especially important to develop the field of asteroseismology for both main-sequence and giant stars to better understand the physics of stellar structure and convective energy transport.

*3/ Extragalactic domain*

Our goal will be to motivate use of the ultimate power of telescopes and interferometry on the Moon to help unlock the mysteries of our cosmic beginnings, notably inflation, and to map out the first objects in the Universe. Other goals in cosmology will be to detect the first stars, galaxies, and active galactic nuclei at the end of the Dark Ages, and with an Earth-Moon Intensity Interferometer we will have the capability to resolve the images of lensed quasars.

*4/ Communication with interstellar probes*

After 2050, projects for interstellar missions to explore in situ the nearest exoplanets will be well advanced. One of the problems will be how to retrieve information from such a mission. A possible solution is to code the data by bistable metre-wide screens, flipping from opaque to transparent at a frequency of several Hertz . When in transit across the parent star of the planet, they will block the stellar light when in the opaque position. With a resolution of 60 m at Proxima Centauri, the resulting dip in the signal would be ~ 1 %, detectable with an Earth-Moon Intensity Interferometer.

These topics are summarized in Table 1.

| **Science Domain** | **Characteristics to be explored** | **Technique** |
|---|---|---|
| Exoplanets | Exoplanet period of rotation, albedo, radius | Photometry |
| | Coarse surface features of exoplanet | Spectro-photometry |
| | Molecular composition and pressure of exoplanet atmosphere | Spectroscopy |
| | Surface composition of rocky exoplanets (with clear skies) | Spectroscopy |
| | Surface structure and features of exoplanets | Intensity Interferometer |
| | Disentangle signals due to exoplanet from stellar speckles | Polarimetry |
| | Biosignature (of vegetation) | Spectroscopy, polarimetry |
| Stellar Physics | Dynamics of stellar atmospheres | Spectroscopy, polarimetry |
| | Stellar interiors | Spectroscopy, polarimetry |
| | Stellar environments, including circumstellar disks and | Spectroscopy, polarimetry |

|  | planetary systems |  |
|---|---|---|
| Cosmology | First stars, galaxies, and Active Galactic Nuclei | Intensity Interferometer |
|  | Microlensing of quasars | Intensity Interferometer |
| Other | Communication link with interstellar probes | Intensity Interferometer |

*Table 1: Science objectives for OWL-Moon.*

## Science capabilities of OWL-Moon

The challenge of imaging an exoplanet is huge but not insurmountable. Consider the following: The Earth is 10 billion times fainter than the Sun and orbits close to its host star. When viewed from 100 pc, the separation is only 0.01 arcsec. This is well above the diffraction limit of a very large lunar telescope, hence it would in principle be possible to study exoplanet atmospheres from a lunar platform, where there is no atmosphere to confuse the signal.

We compare the capabilities of OWL-Moon (a 50-100m telescope) and LUVOIR (a 15m space telescope) to detect Earth-like exoplanets (Sandora and Silk 2020). LUVOIR is capable of observing exoplanet spectra of about 50 Earth-like planets around main sequence stars within 25 pc over the course of a 25-year mission (LUVOIR interim study, 2018). Applying the rule of scaling with telescope diameter (Stark et al. 2014), we find for noise dominated by zodiac light that 3000 planets could be imaged by OWL-Moon, while for photon count-limited noise the number of exoplanets that OWL-Moon could image is 1000. Thus OWL-Moon has significantly enhanced chances compared to LUVOIR of finding atmospheric biosignatures for exoplanets within a search radius of 25 pc

Another strong benefit of Intensity Interferometry will be in the extragalactic domain, namely 2-D images of microlensing. As an example, let us consider high angular resolution of microlensed quasars. If a point source at a distance $D_S$ is perfectly aligned with a foreground source at a distance $D_L$ with a mass $M_L$, the latter acts as a gravitational lens and creates an annular image (Einstein ring) of the background source with an angular radius

$$\theta_E = \sqrt{\left(\frac{4GM_L}{c^2}\right)\frac{(D_S - D_L)D_L}{D_L}}$$

and an angular thickness $\theta_S/2$ where $\theta_S$ is the angular diameter of the source in the absence of gravitational lensing. When the foreground lens is off-axis with respect to the lensed source, the Einstein ring breaks into a large and a small arc (see Fig 1).

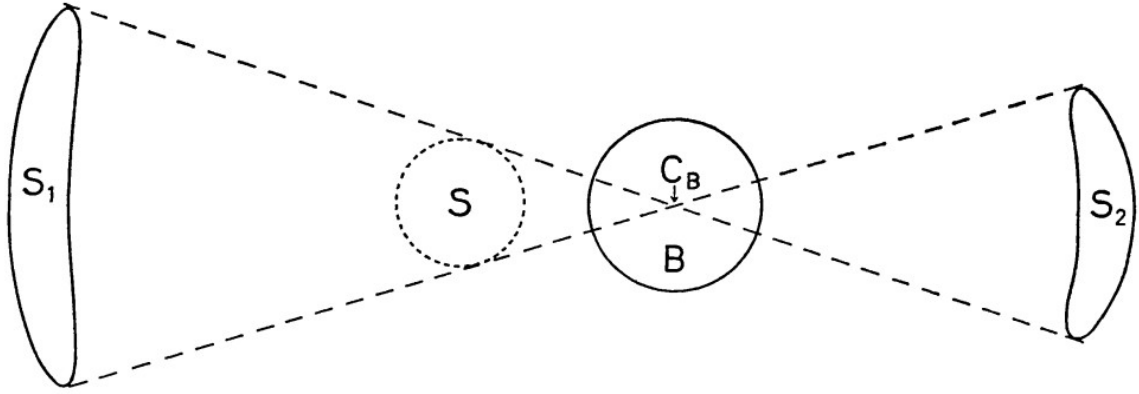

Fig. 1: An off-axis foreground lens C breaks the image of the background source S into a large and a small arc whose images are indicated by S1 and S2. The sizes of S1 and S2 and their angular separation depend on the source and lens distances, on their angular separation and on the lens mass (Refsdal 1964).

A stellar-type microlensing event has too short a lifetime for its detection by Intensity Interferometry but extragalactic, more specifically quasar, microlensing lasts for at least some decades. Indeed, consider the example of the twin quasar QSO 0957+561. The Galaxy mass and distance are respectively $M_L = 8.7 \: 10^{11} M_\odot$ (Vanderriest et al. 1989) and $D_L = 1.6$ Gpc (Kundic et al. 1997). The lensing equation does not include the quasar mass which can therefore be ignored. With a redshift $z = 1.4$ (Walsh et al. 1979), for a Hubble Constant $H_o = 70$ km/s/Mpc its proper distance is S = 5 Gpc. With for instance a relative transverse velocity of say 1,000 km/s for the Galaxy and the quasar, the angular separation between the lensing galaxy and the quasar will change by $1.3 \: 10^{-7}$ arcsec/year, compared to the present 6 arcsec separation between the two images of the quasar. Therefore the configuration of the quasar and galaxy images will last for at least several decades. From the ratio of 3 of the image brightness (Vanderriest et al. 1989) one infers that the ratio of the length of the two arcs is 3. One could approach the central supermassive black hole via resolving broad line emission regions, at less than 100 µarcsec angular size, corresponding typically to a scale of 0.2 pc (Sturm et al. 2018). In addition, one could map the complex structures of quasar narrow emission line regions at arcsecond scales, resolved at ~100 pc scales in nearby AGN.

**Implementation**

The OWL-Moon concept could benefit from the infrastructure afforded by the ESA Moon Village initiative (Crawford 2017) as well as many technological advances that will have been made in the lead up to the Voyage 2050 period.

Current examples of the direction of technology development include the Nautilus project (Apai et al. 2019), which has designed new technology for cheap and light 8m-class telescopes, or the Wide Aperture Exoplanet Telescope project (Monreal et al. 2019), which aims at large (2m 100m in an early prototype), low-cost telescopes. Although the optical quality of these two projects would not be suited for standard interferometry, they would suffice for Intensity Interferometry and high-resolution spectroscopy.

A 100m optical aperture could be constructed from an ensemble of off-axis parabolic surfaces. A proof-of-concept phase for this could begin with a smaller prototype configuration, based on synergy with the PLANETS telescope project (http://www.planets.life/), a 1.9m off-axis telescope under construction at Haleakalā, Hawaii.

Before building a full OWL-Moon configuration, several precursors (from 1m to 8m diameter) should be installed on the lunar south pole, to explore early science cases and test Intensity Interferometry for the brightest sources.

## Global context

The OWL-Moon concept fits well into the Moon Village initiative described by ESA's Director General Jan Wörner as being intended to "bring interested parties together so as to achieve at least some degree of coordination and exploitation of potential synergies" (Woerner 2016). Indeed, the major space agencies of China, India, Japan (Jaxa Press Release 2021), and USA are already developing activities on and around the Moon, for example, China's Lunar Exploration Program (the Chang'e project), India's Chandrayaan-3 mission, and NASA's ARTEMIS program.

As an Earth-Moon interferometer, OWL-Moon will have ground-based partners. For example, the European Extremely Large Telescope (E-ELT) will start operations in the mid-2020s and by 2035 it can be expected to be available to serve as one of the ground-based counterparts of the Earth-Moon Intensity Interferometer.

Finally, we note that the potential use of lunar resources in facilitating the construction of scientific infrastructure on the Moon cannot be underestimated. For example, the HST project was greatly facilitated by the human spaceflight programme (National Academy of Sciences 2005). Similar logic is likely to apply to lunar exploration, with lunar infrastructure likely to subsidize otherwise fiscally unachievable projects such as OWL-Moon.

**Data Availability Statement**
There are no data associated with the manuscript.

**Conflict of Interest**
There are no conflict of interest.